\documentclass{llncs}

\usepackage{amsfonts}
\usepackage[dvips]{graphicx}
\usepackage{amsmath}
\usepackage{amssymb}
\usepackage{xspace}
\usepackage{colortbl}
\usepackage{mathtools}
\usepackage[utf8]{inputenc}
\usepackage[english]{babel}
\usepackage{fancyhdr}
\usepackage{tabularx}
\usepackage{rotating}
\usepackage{comment}
\usepackage{color}
\usepackage{esvect}
\usepackage[T1]{fontenc} 
\usepackage{stmaryrd}
\usepackage{xcolor}
\usepackage{hyperref}
\usepackage{tikz}
\usetikzlibrary{quantikz}
\usepackage[ruled,lined]{algorithm2e}
\usepackage{lineno}
\usepackage{color}
\usepackage[normalem]{ulem}

\renewcommand{\leq}{\leqslant}
\renewcommand{\geq}{\geqslant}

\usepackage{url}

\newcommand{\COMMENT}[1]{}

\title{Longest Common Substring and Longest Palindromic Substring in $\tilde{\mathcal{O}}(\sqrt{n})$ Time\thanks{This work is funded by ICSC -- Centro Nazionale di Ricerca in High-Performance Computing, Big Data and Quantum Computing.}}

\author{Domenico Cantone$^{\dag}$, Simone Faro$^{\dag}$, Arianna Pavone$^{\ddag}$ and Caterina Viola$^{\dag}$}
\institute{$^{\dag}$Department of Mathematics and Computer Science, University of Catania\\
\email{\{domenico.cantone,simone.faro,caterina.viola\}@unict.it}\\[0.1cm]
$^{\ddag}$Department of Mathematics and Computer Science, University of Palermo\\
\email{ariannamaria.pavone@unipa.it}
}

\begin{document}

\maketitle

\newcommand{\pBlock}[1]{p\llbracket #1 \rrbracket}

\begin{abstract}
\vspace{-0.5cm}
The Longest Common Substring (LCS) and Longest Palindromic Substring (LPS) are classical problems in computer science, representing fundamental challenges in string processing. 
Both problems can be solved in linear time using a classical  model of computation, by means of very similar algorithms, both relying on the use of suffix trees. 
Very recently, two sublinear algorithms for LCS and LPS in the quantum query model have been presented by Le Gall and Seddighin~\cite{GallS23},  requiring $\tilde{\mathcal{O}}(n^{5/6})$ and $\tilde{\mathcal{O}}(\sqrt{n})$ queries, respectively.
However, while the query model is fascinating from a theoretical standpoint, its practical applicability becomes limited when it comes to crafting algorithms meant for actual execution on real hardware.
In this paper we present, for the first time, a $\tilde{\mathcal{O}}(\sqrt{n})$ quantum algorithm for both LCS and LPS working in the circuit model of computation. 
Our solutions are simpler than previous ones and can be easily translated  into quantum procedures.  We also present actual implementations of the two algorithms as quantum circuits working in $\mathcal{O}(\sqrt{n}\log^5(n))$ and $\mathcal{O}(\sqrt{n}\log^4(n))$ time, respectively.
\end{abstract}

\section{Introduction}


Quantum computing is a rapidly developing field within computer science that utilizes the principles of quantum mechanics to create more powerful computing systems that operate in a markedly different way from classical computers. 
Quantum computing has had a significant impact on the development of algorithms, with some of the most notable advancements being Shor's algorithm~\cite{Shor1997} for factoring large numbers and Grover's algorithm~\cite{Grover96} for unstructured search. These algorithms provide exponential and quadratic speed-ups over classical algorithms, respectively, serving as impressive demonstrations of the power of quantum computing and sparking a surge of interest in further research and development in the field.
However, it is the recent demonstration of quantum supremacy that has unleashed a wave of interest in quantum computing, leading to the integration of these new technologies in various areas of computer science.

Only recently, text processing and string problems have become a topic of interest within the realm of quantum computation (see for instance \cite{RAMESH2003103,Niroula21,CF23}). 
This paper focuses on the fundamental Longest Common Substring (LCS) and Longest Palindromic Substring (LPS) problems, which hold a crucial position in the field of string processing.  
The LCS problem asks for the longest substring that appears in two given input strings $x$ and $y$ of the same length $n$.
The LPS problem aims to determine the longest substring of a given string $x$ of length $n$ that remains unchanged when read forwards or backwards. 

In the realm of classical computation,  
the LCS and LPS problems admit a linear time solution~\cite{cormen01introduction}, which makes use of generalized suffix trees~\cite{323593}. Remarkably, the solutions to these problems are near identical, and involve the construction of the suffix trees for the input strings and the identification of the lowest common ancestors among the tree nodes.
It is reasonable to wonder about harnessing quantum technology to solve LCS and LPS more efficiently. 

In a recent paper~\cite{GallS23},  
Le Gall and Seddighin proposed  several quantum solutions for the LCS and LPS problems, all based on a composition of other known quantum algorithms, such as Grover search~\cite{Grover96}, string matching~\cite{RAMESH2003103}, element distinctness~\cite{Ambainis07}, and amplitude amplification and estimation~\cite{Brassard2002}. The efficiency of their solutions  
is measured in the \emph{query complexity} model~\cite{RAMESH2003103,Montanaro14}, also known as the \emph{quantum oracle model} (see~\cite{Arora_2023}).  
In such a 
scenario, 
the input is presented as a \emph{black box} that can be accessed by an \emph{oracle} that, given a function $f$, returns the image of the input (or other variables depending on it) via $f$. The query complexity of an algorithm expressed in this model is defined as the number of \emph{queries} that the algorithm makes to the oracle(s). 

Specifically, for the LPS problem they proposed a solution requiring $\tilde{\mathcal{O}}(n^{1/2})$ queries,
while for the LCS problem they proposed a solution requiring $\tilde{\mathcal{O}}(n^{5/6})$ queries and a $1-\varepsilon$ approximate solution requiring $\tilde{\mathcal{O}}(n^{3/4})$ queries. We point out that such solutions are only of theoretical interest and can hardly be translated into actual quantum circuits.
For the sake of completeness, we mention also two solutions for the case of \emph{non-repetitive strings}\footnote{A non-repetitive string of length $n$ is a string where no character of the alphabet appears twice. We note that this implies that $n \leq |\Sigma|$.} (see Table \ref{tab:lcs-solutions}).

The query model allows one to study and analyse quantum algorithms without worrying about the technicalities around the construction of any specific oracle.  The query model has been the framework of the first outstanding attempts to design algorithms that exhibit a theoretical advantage against classical ones. 
While the query model is intriguing and abstract enough to be valuable for purely theoretical approaches, it holds limited significance when it comes to designing algorithms intended for practical implementation on real hardware.
Indeed, it is often unclear how one could implement a phase oracle efficiently.

However, there are different models of quantum computation that easily and almost directly translate to concrete implementations on quantum computers.\footnote{We observe that contemporary quantum computers do not have yet the memory capabilities to deal with the large number of qubits involved in our algorithm.} 
Perhaps, one of the most widespread among such models is the \emph{quantum circuit model}~\cite{Yao93}, considering that there are several programming languages featuring the circuit formalism, such as IBM's Qiskit, Microsoft's $\#$Q, and Google's Cirq, just to cite a few. In this context, a quantum circuit can be thought as a Boolean circuit in which the basic gates represent elementary quantum operations, and the computational complexity of a quantum circuit is measured by its \emph{depth}.

In this paper we present the first quantum algorithms for the LCS and LPS problems in the circuit model of computation, providing actual implementations of quantum circuits that work in $\tilde{\mathcal{O}}(\sqrt{n})$ computational time.
Specifically, we define a unique approach that can be adapted to solve both problems. For the LCS, the proposed approach leads to the definition of an effective circuit that requires $\mathcal{O}(\sqrt{n}\log^4(n))$ time in the case of binary strings, and $\mathcal{O}(\sqrt{n}\log^5(n))$ time in the general case. For the LCS, we obtain two circuits working in $\mathcal{O}(\sqrt{n}\log^3(n))$ and $\mathcal{O}(\sqrt{n}\log^4(n))$ time in the two cases, respectively.

Regarding the quantum access to any input string $s$, in~\cite{GallS23} the authors assume a QRAM (Quantum Random Access Memory) model~\cite{phalak2023quantum}.
Specifically, it is assumed that the string $s$ can be accessed directly by a random access oracle that performs, at unit cost, unitary mappings of the kind $|i\rangle |a\rangle |z\rangle \rightarrow |i\rangle |a \otimes s[i]\rangle |z\rangle$, where $i$ is a string position, such that $0\leq i < n$,  $a \in \Sigma$ is a character, $z \in \{0,1\}^*$, and $\otimes$ denotes an appropriate binary operation defined on $\Sigma$.
However, we point out that the most efficient QRAM designs~\cite{Giovannetti_2008,Giovannetti_2008b} exhibit a polylogarithmic time complexity for accessing the memory with respect to its size. In our scenario, the memory size is $\mathcal{O}(n)$, which implies that QRAM queries will incur an additional multiplicative cost of at least $\mathcal{O}(\log^2(n))$. Moreover, we must consider the overhead of initializing the quantum memory, which requires $\mathcal{O}(n)$ operations~\cite{Park_2019}.
In contrast, our algorithm does not rely on a random access oracle, but we offer a detailed circuit for solving the problem, constructed using elementary gates.

Ultimately, our approach stands apart from the previous results due to its inherent simplicity, which enables us not only to provide a circuit-level blueprint, but also to assess the quantum resources required for its implementation.

Finally, we observe that, in~\cite{GallS23}, the authors discovered the unexpected fact that, despite their strong correlation, the quantum computational complexities of the LCS and LPS problem are actually distinct. As evidence of this, they also proved that any quantum algorithm for LCS must take at least $\tilde{\Omega}(n^{2/3})$ time, even when binary strings are considered.
Since we work within a different computational model, the results presented in this paper do not contradict the aforementioned lower bound. However, the fact that our approach can be used to solve both problems with the same computational complexity 
provides additional confirmation of their strong correlation, even in the quantum field.

The paper is organized as follows.  In Section \ref{sec:preliminaries}, we review some useful preliminaries.
Next, in Section \ref{sec:lcs} we provide an abstract view of the algorithm for solving the LCS problem. 
Then, in Section \ref{sec:circuit}, we describe an actual implementation of the same algorithm within the circuit-based model. Finally, in Section \ref{sec:lps}, we briefly show how to adapt our solution to  LPS problem, and draw our conclusions.

\begin{table}[!t]
\begin{center}
\begin{tabular}{|c|c|c|c|c|c|}
	\hline
	~~Problem~~ & ~~Paper~~ & ~~Case~~ & ~~Query Compl.~~ & ~~Time Compl.~~\\
	\hline
	~~exact LPS~~ &  				~\cite{GallS23} & general case  & $\tilde{\mathcal{O}}(\sqrt{n})$ 	&  - \\
	~~exact LCS~~ &  				~\cite{GallS23} & general case  & $\tilde{\mathcal{O}}(n^{5/6})$ 	&  - \\
	~~exact LCS~~ &  				~\cite{GallS23} & ~non-repetitive~ & $\tilde{\mathcal{O}}(n^{3/4})$ 	&  - \\
	~~exact LPS~~ &  				 ~This paper~      & general case  & - 	&  ~$\tilde{\mathcal{O}}(\sqrt{n})$~ \\
	~~exact LCS~~ &  				 ~This paper~      & general case  & - 	&  $\tilde{\mathcal{O}}(\sqrt{n})$ \\
	\hline
\end{tabular}
\end{center}
\caption{\label{tab:lcs-solutions}A comparison of quantum solutions on LPS and LCS. 
The time complexities for \cite{GallS23} are unknown due to their reliance on random-access oracles, which lack a circuit-level construction in the referenced paper.}
\end{table}

\section{Preliminaries}\label{sec:preliminaries}

We represent a string $x$ of length $n \geq 1$, over a finite alphabet $\Sigma$ of size $\sigma$, as a finite array $x[0\,..\,n-1]$, 
and denote the empty string by $\varepsilon$.
We also denote by $x[i]$ the $(i+1)$-st character of $x$, for $0\leq i< n$, and by $x[i\,..\,j]$ the substring of $x$ contained between the $(i+1)$-st and the $(j+1)$-st characters of $x$, for $0\leq i \leq j <n$.  
A $k$-substring of a string $x$ is any substring of $x$ of length $k$. 
For ease of notation, the $(i+1)$-st character of the string $x$ will also be denoted by the symbol $x_i$, so that $x = x_0 x_1 \ldots x_{n-1}$.   
A substring of $x$ beginning at position $0$ is a \emph{prefix} of $x$.  We use the notation $x_{:i}$ to indicate the prefix of $x$ of length $i$.

For any two strings $x$ and $y$ of length $n$, we say that $x$ and $y$ have a common $k$-substring  if there exist two indices $0\leq i,j<n-k$ such that $x[i \,..\, i+k-1] = y[j \,..\, j+k-1]$. In particular, when the indices $i$ and $j$ coincide, we say that $x$ and $y$ \emph{share} a $k$-substring at position $i$. The expression $x\cdot y$ denotes the concatenation of $x$ and $y$. Furthermore, given a string $x$ of length $n$ and a shift $0\leq j <n$, we denote by $\vv{x}^j$ the cyclic rightward rotation of the characters of $x$ by $j$ positions. More formally, we have $\vv{x}^j \coloneqq x[n-j \,..\, n-1]\cdot x[0 \,..\, n-j-1]$.

Due to  space limitations, we assume the reader's familiarity with essential concepts in quantum computation, including qubits, bra-ket notation, amplitudes, quantum entanglement, and measurement. %
Multiple qubits taken together are referred to as \emph{quantum registers}. Specifically, a quantum register $|\psi\rangle = |q_0, q_1, \ldots, q_{n-1}\rangle$ of $n$ qubits is the tensor product $\bigotimes_{i=0}^{n-1}|q_i\rangle$ of its constituent qubits.
If $k$ is an integer value that can be represented as binary string of length $n$, we use the symbol $|k\rangle$ to denote the register $\bigotimes_{i=0}^{n-1}|k_i\rangle$ of $n$ qubits, where $|k_i\rangle$ takes the value of the $i$-th most significant binary digit of $k$. 
Thus, the quantum register $|8\rangle$ with $4$ qubits is given by $|8\rangle = 
|1000\rangle$.

\emph{Operators} in quantum computing are mathematical entities used to represent functional processes that result in the change of the state of a quantum register. 
Although there is no problem in realizing any quantum operator capable of working in constant time on a quantum register of fixed size, operators of variable size can only be implemented through the composition of elementary gates.

Given a function $f\colon \{0,1\}^n \rightarrow \{0,1\}$, any quantum operator that  maps a register containing the value of a given input $x \in \{0,1\}^n$ into a register whose value depends on $f(x)$ is called a \emph{quantum oracle}.  
A \emph{Boolean oracle} $U_f$ maps a register $|x\rangle \otimes |0\rangle$, of size $n+1$, to the register $|x\rangle \otimes |f(x)\rangle$. More formally,  $U_f |x,0\rangle = |x,f(x)\rangle$.
A \emph{phase oracle} $P_f$ for a function $f\colon \{0,1\}^n \rightarrow \{0,1\}$ takes as input a quantum register $|x\rangle$, where $x\in \{0,1\}^n$, and leaves its value unchanged, while applying to it a negative global phase only when $f(x)=1$, that is, only if $x$ is a solution for the function. More formally,  $P_f |x\rangle = (-1)^{f(x)}|x\rangle$.

We adopt the circuit model of computation, and we assume any circuit as being divided into a sequence of discrete time-steps, where the application of a single gate requires a single time-step.  In this paper, we adopt the \emph{depth} of the circuit as our chosen measure for time complexity, which refers to the total number of required time-steps. We remark that the depth of a circuit is not necessarily the same as the total number of gates (called \emph{size}) of the circuit itself, since gates that act on disjoint qubits can often be applied in parallel.

\section{Quantum Longest Common Substring}\label{sec:lcs}
 
In this section, we first describe our quantum algorithm for computing the LCS within an abstract model, 
in order to better understand its design, by defining the quantum oracles involved in the computation, but without giving their actual implementation. Later, we will show that our abstract algorithm 
requires $\tilde{\mathcal{O}}(\sqrt{n})$ queries to the oracles.\footnote{We would like to point out that in counting the number of queries requested by our algorithm, we do not intend to compute its query complexity, since we work within the circuit-based computational approach that does not conform to the constraints of the query-based model.}
Subsequently, we will present an actual implementation of our algorithm in the circuit-based computational model.

Our algorithm, named \textsc{Quantum-LCS}, comprises a quantum computation-based and a classical computation-based components. Its simple underlying structure is summarized in the pseudocode shown in Fig. \ref{fig:algo}. 

The classical part of the computation consists of a binary search for identifying the length $d$ of the longest common substring of $x$ and $y$ (line $2$). 

During each iteration, we check for the presence of a common substring of length $d$ between $x$ and $y$.
Let $[\ell \,..\, r]$ be the interval over which the binary search is restricted during an iteration of the algorithm, and let $d=\lfloor (\ell + r) / 2 \rfloor$ be its median. The values of $\ell$ and $r$ are initialized to $0$ and $n$, respectively. At each iteration, it checks for the presence of a common substring of length $d$ between $x$ and $y$. 
If the iterative test returns a positive answer, then the interval is narrowed to $[d \,..\, r]$, otherwise it is narrowed to $[\ell \,..\, d-1 ]$. 
The search identifies the length $d$ of the longest common substring in $\mathcal{O}(\log(n))$ steps.

The quantum part of the algorithm implements the test of line $4$.  
In what follows we will focus exclusively on the implementation of such iterative test.

Before describing the details of the quantum procedure for the iterative test, it is necessary to formalize some assumptions we make along the description.

Since a quantum register of dimension $\log(n)$ can take on all values between $0$ and $n-1$, like any binary sequence of the same dimension, for simplicity we will assume that both input strings $x$ and $y$ have length $n=2^p$, for some $p>0$. We also assume that $x$ and $y$ end with two different special characters, \$ and \%, respectively, not belonging to the alphabet $\Sigma$. These assumptions can be made without any loss of generality, since it would suffice to take the smallest value $p$ for which we have $n<2^p$ and concatenate the text with $2^p-n$ copies of the special character. 
For instance, if $x=\texttt{abaacbcbbca}$ is a text of length $11$, we silently concatenate it with $5$ copies of the character \$, i.e., we assume that $x = \texttt{abaacbcbbca\$\$\$\$\$}$. This assumption has no effect on the complexity, since the resulting string has at most twice the length of the original string.

\begin{figure}[!t]
\begin{footnotesize}
\begin{tabular}{lll}
	\multicolumn{3}{l}{\textsc{Quantum-LCS}$(x,y,n)$:}\\
	1. & \textsf{$\ell \leftarrow 0$; $r \leftarrow n$} &\\
    	2. & \textsf{while $\ell < r$ do} &\\
        	3. & \quad \textsf{$d \leftarrow \lfloor (\ell + r) / 2 \rfloor$} &\\
        	4. & \quad \textsf{if \fbox{$\exists\ i,j\in \{0,\ldots,n-1\}\ :\  \vv{x}^j[i\, ..\, i+d-1] = y[j\, ..\, j+d-1]$}} &$\leftarrow$\ \textsc{Quantum test}\\
        	5. & \quad \quad \textsf{then $\ell \leftarrow d$} &\\
        	6. & \quad \quad \textsf{else $r \leftarrow d-1$} &\\
	7. & \textsf{return $\ell$} &
\end{tabular}\\[-0.3cm]
\end{footnotesize}
\caption{\label{fig:algo}The pseudocode of the algorithm for computing the LCS between two strings, $x$ and $y$, of length $n$. The quantum part of the algorithm reduces to the iterative test of line 4.} 
\end{figure}

\smallskip

Despite any substring of length $d$ can begin at any position $j$ of the text, for $0\leq j\leq n-d$, in this paper we also admit values of $j$ between $0$ and $n-1$, thus assuming that a substring of the text can be obtained in a circular way. Even such an  assumption can be made without loss of generality, since the last character of $x$ and $y$ are the special character \$ and \%, respectively, and therefore no substring obtained circularly can ever be returned as LCS.

For the sake of simplicity and due to space constraints,  we restrict to circuits algorithms designed for processing binary strings. This further simplification, however, does not lead to any substantial change in our results since, assuming that each character can be represented with (at most) $\log(n)$ bits, it is easy to show that the quantum operators used in the construction of the algorithm would undergo an increase in their complexity at most equal to a factor of $\log(n)$.

\subsection{The Quantum Iterative Test}\label{sec:quantum_lcs}

Given two strings $x$ and $y$, both of length $n$, and a bound $d\leq n$, the quantum test checks for the presence of a common substring of length $d$ between $x$ and $y$. 

The abstract procedure that implements the iterative test is schematized in Fig.~\ref{fig:quantum-test}. It consists of three phases, each of which implemented by a quantum sub-procedure: (1) a search phase, (2) a verification phase, and (3) a final check. The output of the iterative test is the output of the final check. In this section we describe in detail the role, structure, and complexity of each of the phases and then discuss the overall complexity of the iterative test.

\smallskip

The \emph{search phase} makes use of the Grover's search algorithm~\cite{Grover96} 
for unstructured search for finding (with high probability) a solution (if any) to a black box function, making just $\mathcal{O}(\sqrt{n})$ queries to the function.

Specifically, the input black box to the algorithm is accessed by a phase oracle $P_{\psi}$ implementing the function $\psi_{(x,y)} \colon \{0,\ldots,n-1\} \times \{0,\ldots,n-1\} \longrightarrow \{0,1\}$, which depends on the strings $x$ and $y$. Given the two input parameters $j$ and $d$, with $0\leq j,d<n$, the phase oracle $P_{\psi}$  tests whether the two strings $\vv{x}^j$ and $y$ share a $d$-substring. The function $\psi_{(x,y)}$ is defined, for all $0\leq j,d< n$, as
\begin{small}
\begin{equation}\label{eq:f}
	\psi_{(x,y)}(j,d) = \left\{ \begin{array}{ll}
		1 & \textrm{ if } \exists\ i \in \{0,\ldots,n-1\}:\vv{x}^j[i \,..\, i+d-1] = y[i \,..\, i+d-1]\\
		0 & \textrm{ otherwise.}
	\end{array}
	\right.
\end{equation}
\end{small}

When the values of $x$ and $y$ are clear from the context, for simplicity we will use the symbol $\psi$ instead of $\psi_{(x,y)}$.

Using this convention, the phase oracle $P_{\psi}$ operates so as to achieve the transformation
$
	P_{\psi} |j \rangle  = (-1)^{\psi(j)} |j \rangle
$, for all $j \in \{0,1\}^{\log(n)}$.\footnote{In section \ref{sec:circuit-search} we show how the phase oracle $P_{\psi}$ can be effectively implemented by means of a circuit having depth $\mathcal{O}(\log^3(n))$.}

After $\mathcal{O}(\sqrt{n})$ iterations of Grover's algorithm, the procedure returns a potential solution $j$ to the problem, such that  $\vv{x}^j$ and $y$ share a $d$-substring. 
However, since such a solution may not exist (a case in which the search would return a random state $0\leq j <n$), it is necessary to run the subsequent verification procedures to check whether the returned state is an actual solution of the function.

\smallskip

Assuming $\vv{x}^j$ and $y$ share a $d$-substring, the \emph{verification phase} again makes use of Grover's search algorithm in order to identify a position $i$ within the strings such that $\vv{x}^j[i \,..\, i+d-1] = y[i \,..\, i+d-1]$. 
In this case, the input black box to the algorithm is a phase oracle $P_{\varphi}$ implementing the function $\varphi_{(x,y)} \colon \{0,\ldots,n-1\} \times  \{0,\ldots,n-1\} \times  \{0,\ldots,n-1\} \longrightarrow \{0,1\}$, where, for all strings $x$ and $y$,  and for all $0\leq i,j,d <n$, we have:
\begin{equation}\label{eq:t}
	\varphi_{(x,y)}(i,j,d) = \left\{ \begin{array}{ll}
		1 & \textrm{ if } \vv{x}^j[i \,..\, i+d-1] = y[i \,..\, i+d-1]\\
		0 & \textrm{ otherwise.}
	\end{array}
	\right.
\end{equation}
Therefore, the oracle $P_{\varphi}$ operates so as to achieve the phase transformation 
$
	P_{\varphi} |i \rangle |j \rangle |d \rangle  = (-1)^{\varphi(i,j,d)} |i \rangle |j \rangle
$, for all $i,j,d \in \{0,1\}^{\log(n)}$.\footnote{In section \ref{sec:circuit-search} we give an implementation of the phase oracle $P_{\varphi}$ in $\mathcal{O}(\log^3(n))$ time.}

Even in this case, after $\mathcal{O}(\sqrt{n})$ iterations of Grover's algorithm, the procedure returns a potential position $i$, such that  $\vv{x}^j[i \,..\, i+d-1] = y[i \,..\, i+d-1]$. 

Ultimately, the quantum test ends by checking whether the two substrings of length $d$ beginning at position $i$ of the strings $\vv{x}^j$ and $y$ are indeed equal. Such a \emph{final check} can be exactly computed through a single execution of the quantum oracle $U_{\varphi}$ implementing the function $\varphi$ defined in (\ref{eq:t}).

\begin{figure}[!t]
\begin{footnotesize}
\begin{tabular}{lll}
	\multicolumn{3}{l}{\textsc{Quantum test}:}\\
        	1. & \textsf{$j \leftarrow$ get a random $k$ such that $\vv{x}^k$ and  $y$ share a $d$-substring} &~~(\textsc{Search phase})\\
        	2. & \textsf{$i \leftarrow$ get a random $k$ such that $\vv{x}^j[k\, ..\, k+d-1] = y[k\, ..\, k+d-1]$} &~~(\textsc{Verification})\\
        	3. & \textsf{check if $\vv{x}^j[i\, ..\, i+d-1]$ is equal to $y[i\, ..\, i+d-1]$} &~~(\textsc{Final check})\\
\end{tabular}\\[-0.3cm]
\end{footnotesize}
\caption{\label{fig:quantum-test}The structure of the quantum test used in the algorithm in Fig.\ref{fig:algo}. The test consists of 3 phases, each of which is implemented as a quantum procedure.} 
\end{figure}

The whole structure of the quantum test is depicted in Fig.\ref{fig:quantum-test}. The first two phases require both $\mathcal{O}(\sqrt{n})$ queries to the oracles $P_{\psi}$ and $P_{\varphi}$, respectively, while the last check requires a single query to the boolean oracle $U_{\varphi}$. 
Therefore, the quantum iterative test requires $\mathcal{O}(\sqrt{n})$ queries to quantum oracles.

We point out that, when a solution exists, both the search and verification phases may fail with a probability $\mathcal{O}(1/n)$, due to the internal randomness of Grover's algorithm. When the search phase or the verification phase returns a value that is not a solution of the respective function, the final check fails by returning the value 0. In such a case, we can simply repeat the whole test an arbitrary constant number of times in order to suppress the probability of failure. The test terminates when a common substring between $x$ and $y$ is found, or alternatively, when such an attempt fails an arbitrary number of times. 

The overall number of queries needed to solve the problem is $\mathcal{O}(\sqrt{n}\log(n))$, since the execution of the quantum iterative test requires $\mathcal{O}(\sqrt{n})$ queries and the binary search for the length of the LCS requires $\mathcal{O}(\log(n))$  iterations.


\section{A Circuit-Model Based Implementation}\label{sec:circuit}
In this section we provide an actual implementation of the iterative test shown in Fig.\ref{fig:quantum-test} within the circuit-based computational model. The purpose of this translation is to provide a direct implementation of the algorithm in a quantum computer and evaluate the actual resources required.

The three steps of the iterative test are implemented by the three circuits reported in Fig.\ref{fig:circuit_lcs}. Only three operators are used as building-blocks in the three circuits: the circular shift (ROT) operator, the shared fixed substring checking (SFC) operator, and the fixed prefix matching~(FPM) operator (see Table~\ref{tab:operators}). For lack of space, in this section we only provide a brief overview of how these operators are structured, referring the reader to the appropriate references.\footnote{We observe that the oracles used in the actual circuits of Fig.\ref{fig:circuit_lcs} are implemented as boolean oracles rather than as a phase oracles. These are denoted as $U_{\psi}$ and $U_{\varphi}$ instead of $P_{\psi}$ and $P_{\varphi}$, respectively. However, we recall that initializing the output register of a boolean oracle to the value $|-\rangle$ allows it to behave like a phase oracle.}

\smallskip

A \emph{circular shift operator} (or rotation operator) ROT applies a rightward shift of $s$ positions to a register of $n$ qubits for a fixed parameter $0\leq s <n$. Thus, the element at position $i$ is moved to position $(i+s) \mod n$. 
Such an operator has been effectively used in other quantum text searching algorithms~\cite{Niroula21,CF23}.
The details of its construction have been detailed by Pavone and Viola in~\cite{PV23}, where it is shown that the resulting operator can be executed in $\mathcal{O}(\log(n))$ time, both in the case of binary strings and in the general case.

In our implementation we make use of the controlled version of the circular shift operator, which applies a circular rotation of a number of positions, depending on an input value $j$ such that $0\leq j <n$. 

More formally, for all $x\in \{0,1\}^n$ and all $j \in \{0,1\}^{\log(n)}$, the controlled circular shift operator performs the following mapping:

$$
	\textsc{ROT} |j\rangle |x\rangle = |j\rangle |\vv{x}^j\rangle.
$$

The controlled variant of the circular shift operator can be implemented by means of a well-known technique~\cite{CF23} 
that involves the use of $\log(n)$ ancill\ae\ qubits for the application of all parallel operators controlled by the same qubit, with an  overhead of $\mathcal{O}(\log(n))$.
Thus, the operator can be executed in $\mathcal{O}(\log^2(n))$ time.

\smallskip

\begin{table}[!t]
\begin{footnotesize}
\begin{center}
\begin{tabular}{lllll}
	\textsc{Operator} & ~~\textsc{Symbol}  & ~~\textsc{Binary String} & ~~\textsc{General Case} & ~~\textsc{Ref.} \\
	\hline\\[-0.2cm]
	Controlled Circular Shift 			& ~~ROT & ~~$\mathcal{O}(\log^2(n))$ & ~~$\mathcal{O}(\log^2(n))$ &  ~~\cite{PV23} \\
	Shared Fixed Substring Check 	& ~~SFC & ~~$\mathcal{O}(\log^3(n))$ & ~~$\mathcal{O}(\log^4(n))$ &  ~~\cite{CFPV23} \\
	Fixed Prefix Matching 			&~~FPM & ~~$\mathcal{O}(\log^3(n))$ & ~~$\mathcal{O}(\log^4(n))$ &  ~~\cite{CFPV23} \\
	\hline
\end{tabular}
\end{center}
\end{footnotesize}
\caption{\label{tab:operators}The three operators used in the implementation of our circuits, with an indication of their computational complexity in the case of binary strings and in the general case.}
\end{table}

The \emph{shared fixed substring checking} (SFC) operator addresses the following simple string matching problem, in which,  
given two strings $x$ and $y$, both of length $n$, and a bound $d\geq 0$,  one wants to check whether $x$ and $y$ share a common $d$-substring, i.e., if there exists a position $i$, with $0\leq i< n-d$, such that $x[i\,..\,i+d-1] = y[i\,..\,i+d-1]$. 
In other words, the SFC operator computes the function $\psi_{(x,y)}(j,d)$  for the special case where $j=0$, that is, $x$ does not undergo any cyclic rotation.
More formally, given a bound $d\leq n$, the SFC operator, for all $x,y\in \{0,1\}^n$ and $d\in \{0,1\}^{\log(n)}$, is defined by
$$
	\textrm{SFC} |x\rangle |y\rangle |d\rangle |0\rangle = |x\rangle |y\rangle |d\rangle |\psi_{(x,y)}(0,d)\rangle. 
$$

The construction of a quantum circuit implementing the SFC operator has been recently proposed in \cite{CFPV23}, where the authors provide a circuit with a $\mathcal{O}(\log^3(n))$ depth in the case of binary input strings and a circuit with a $\mathcal{O}(\log^4(n))$ depth in the general case. We do not give here further details on the construction of the operator but refer to \cite{CFPV23}  for any structural aspects of the corresponding circuit.

\smallskip

Given two strings $x$ and $y$, both of length $n$, and a bound $d\leq n$, the \emph{fixed prefix matching} (FPM) operator performs a simple check to determine if the first $d$ characters of the string $x$ match their counterparts in the string $y$. Roughly speaking, the FPM operator checks if $x_{:d}=y_{:d}$.
More formally, for all $x,y\in \{0,1\}^n$, and all $d \in \{0,1\}^{\log(n)}$, the FPM operator is defined by
$$
	\textsc{FPM} |x\rangle |y\rangle |d\rangle |0\rangle = |x\rangle |y\rangle |d\rangle |\varphi_{(x,y)}(0,0,d)\rangle. 
$$

The construction of a quantum circuit implementing the FPM operator has also been recently proposed in \cite{CFPV23}, where the authors give a circuit with a depth of $\mathcal{O}(\log^3(n))$ for the case of binary input strings and a circuit with a depth of $\mathcal{O}(\log^4(n))$ for the general case.

\smallskip

We are now ready to describe the quantum circuits that implement the three phases of the quantum iterative test.
All circuits make use of two registers $|x\rangle$ and $|y\rangle$, both of size $n$, which we assume to contain the characters of the two input strings $x$ and $y$, respectively. We also assume that these registers are already stored in a quantum memory and do not need initialization.
All circuits involve the presence of an input register $|d\rangle$, of size $\lceil \log(d)\rceil \leq \log(n)$, containing the binary representation of the bound $d\leq n$. The initialization of such input register requires $\mathcal{O}(\log(n))$ time.
The output of the computation, for all circuits, is stored in the $|out\rangle$ register consisting of a single qubit.

\subsection{Implementing the Circuits for the Three Phases}\label{sec:circuit-search}
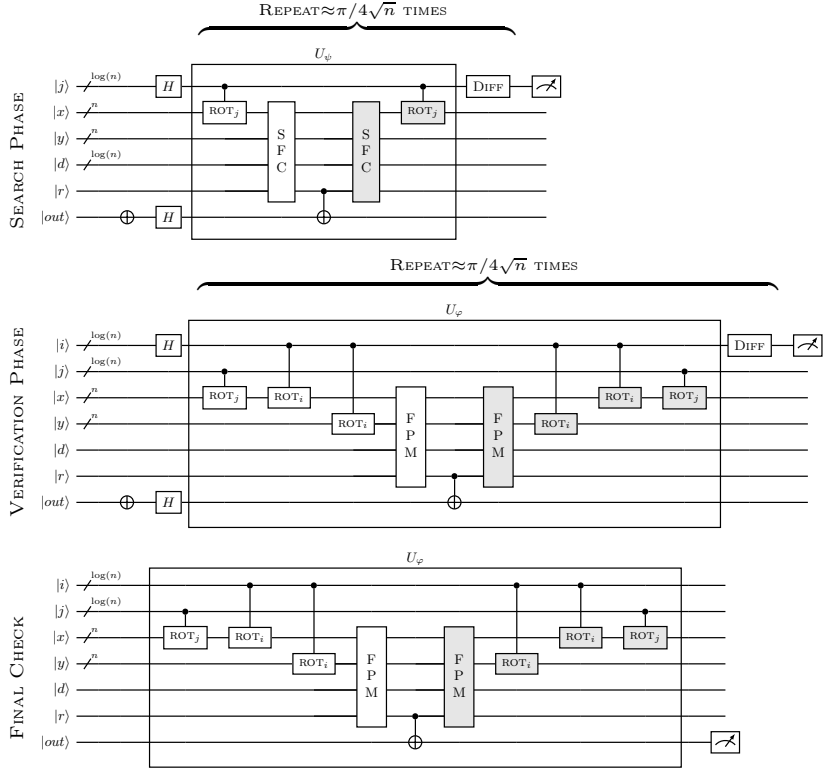
\begin{figure}[!t]
\hspace{2.5cm}$\overbrace{\hspace{4.2cm}}^{\textsc{Repeat} \approx \pi/4 \sqrt{n} \textsc{ times }}$\\
\rotatebox{90}{\scriptsize\textsc{~~Search Phase}}
\begin{tikzpicture}
\node[scale=0.58] {
\begin{quantikz}[row sep={0.6cm,between origins}]
\lstick{\ket{j}} & \qwbundle{\log(n)} 	& \qw & \gate{H} & \ctrl{1}\gategroup[6,steps=5,style={inner sep=4pt}]{$U_{\psi}$} & \qw & \qw & \qw & \ctrl{1} &  \gate{\textsc{Diff}} & \meter{} \\
\lstick{\ket{x}} & \qwbundle{n} 		& \qw & \qw 	 & \gate{\textsc{rot}_j} & \gate[4,disable auto height]{\begin{array}{c} \text{S} \\ \text{F} \\ \text{C} \end{array}} & \qw & \gate[4,disable auto height, style={fill=gray!20}]{\begin{array}{c} \text{S} \\ \text{F} \\ \text{C} \end{array}} &  \gate[style={fill=gray!20}]{\textsc{rot}_j} & \qw & \qw\\
\lstick{\ket{y}} & \qwbundle{n} 		& \qw & \qw 	& \qw & \qw & \qw & \qw & \qw & \qw & \qw\\
\lstick{\ket{d}} & \qwbundle{\log(n)} 	& \qw & \qw 	& \qw & \qw & \qw & \qw & \qw & \qw & \qw\\
\lstick{\ket{r}} & \qw 				& \qw & \qw 	& \qw & \qw & \ctrl{1} & \qw & \qw & \qw & \qw\\
\lstick{\ket{out}} & \qw 			& \targ{} & \gate{H} 	& \qw & \qw & \targ{} & \qw & \qw & \qw & \qw
\end{quantikz}
};
\end{tikzpicture}\\[0.1cm]
{\color{white}.}\hspace{2.4cm}$\overbrace{\hspace{7.7cm}}^{\textsc{Repeat} \approx \pi/4 \sqrt{n} \textsc{ times }}$\\
\rotatebox{90}{\scriptsize\textsc{~~Verification Phase}}
\begin{tikzpicture}
\node[scale=0.58] {
\begin{quantikz}[row sep={0.6cm,between origins}]
\lstick{\ket{i}} & \qwbundle{\log(n)} 	& \qw & \gate{H} & \qw\gategroup[7,steps=9,style={inner sep=6pt}]{$U_{\varphi}$} & \ctrl{2} & \ctrl{3} & \qw & \qw & \qw & \ctrl{3} & \ctrl{2} & \qw  & \gate{\textsc{Diff}} & \meter{} \\
\lstick{\ket{j}} & \qwbundle{\log(n)} 	& \qw & \qw & \ctrl{1} & \qw & \qw & \qw & \qw & \qw & \qw & \qw & \ctrl{1} & \qw & \qw\\
\lstick{\ket{x}} & \qwbundle{n} 		& \qw & \qw 	& \gate{\textsc{rot}_j}  & \gate{\textsc{rot}_i} & \qw & \gate[4,disable auto height]{\begin{array}{c} \text{F} \\ \text{P} \\ \text{M} \end{array}} & \qw & \gate[4,disable auto height, style={fill=gray!20}]{\begin{array}{c} \text{F} \\ \text{P} \\ \text{M} \end{array}} & \qw &  \gate[style={fill=gray!20}]{\textsc{rot}_i} & \gate[style={fill=gray!20}]{\textsc{rot}_j} & \qw & \qw\\
\lstick{\ket{y}} & \qwbundle{n} 		& \qw & \qw & \qw 	& \qw & \gate{\textsc{rot}_i}  & \qw & \qw & \qw & \gate[style={fill=gray!20}]{\textsc{rot}_i} & \qw & \qw & \qw & \qw\\
\lstick{\ket{d}} & \qw 			& \qw & \qw 	& \qw & \qw & \qw & \qw & \qw & \qw & \qw & \qw & \qw & \qw & \qw \\
\lstick{\ket{r}} 	& \qw 			& \qw & \qw	& \qw & \qw & \qw & \qw & \ctrl{1}  & \qw & \qw & \qw & \qw & \qw & \qw\\
\lstick{\ket{out}} & \qw 			& \targ{} & \gate{H} 	& \qw & \qw & \qw & \qw & \targ{} & \qw & \qw & \qw & \qw & \qw & \qw
\end{quantikz}
};
\end{tikzpicture}\\[0.1cm]
\rotatebox{90}{\scriptsize\textsc{~~~~Final Check}}
\begin{tikzpicture}
\node[scale=0.58] {
\begin{quantikz}[row sep={0.6cm,between origins}]
\lstick{\ket{i}} & \qwbundle{\log(n)} 	& \qw & \qw & \qw\gategroup[7,steps=9,style={inner sep=6pt}]{$U_{\varphi}$} & \ctrl{2} & \ctrl{3} & \qw & \qw & \qw & \ctrl{3} & \ctrl{2} & \qw  & \qw & \qw \\
\lstick{\ket{j}} & \qwbundle{\log(n)} 	& \qw & \qw & \ctrl{1} & \qw & \qw & \qw & \qw & \qw & \qw & \qw & \ctrl{1} & \qw & \qw \\
\lstick{\ket{x}} & \qwbundle{n} 		& \qw & \qw 	& \gate{\textsc{rot}_j}  & \gate{\textsc{rot}_i} & \qw & \gate[4,disable auto height]{\begin{array}{c} \text{F} \\ \text{P} \\ \text{M} \end{array}} & \qw & \gate[4,disable auto height, style={fill=gray!20}]{\begin{array}{c} \text{F} \\ \text{P} \\ \text{M} \end{array}} & \qw &  \gate[style={fill=gray!20}]{\textsc{rot}_i} & \gate[style={fill=gray!20}]{\textsc{rot}_j} & \qw & \qw \\
\lstick{\ket{y}} & \qwbundle{n} 		& \qw & \qw & \qw 	& \qw & \gate{\textsc{rot}_i}  & \qw & \qw & \qw & \gate[style={fill=gray!20}]{\textsc{rot}_i} & \qw & \qw & \qw & \qw \\
\lstick{\ket{d}} & \qw 			& \qw & \qw	& \qw & \qw & \qw & \qw & \qw& \qw & \qw & \qw & \qw & \qw & \qw \\
\lstick{\ket{r}} 	& \qw 			& \qw & \qw	& \qw & \qw & \qw & \qw & \ctrl{1}  & \qw & \qw & \qw & \qw & \qw & \qw \\
\lstick{\ket{out}} & \qw 			& \qw & \qw	& \qw & \qw & \qw & \qw & \targ{} & \qw & \qw & \qw & \qw & \qw & \meter{}
\end{quantikz}};
\end{tikzpicture}\\[-0.7cm]
\caption{\label{fig:circuit_lcs}The quantum circuits implementing the three phases of the iterative test of the \textsc{Quantum-LCS} algorithm. 
For simplicity, the circuits do not contain all the ancill\ae\ qubits. The operators in gray color represent the inverse operators.}
\end{figure}

The circuit for the search phase is depicted on the top of Fig.\ref{fig:circuit_lcs}. It makes use of the additional $|j\rangle$ register, of size $\log(n)$, which holds the rotation values of the string $x$. It is initialized to $|+\rangle^{\log(n)}$, in order to maintain, at the initial stage, the superposition of all possible rotation values between $0$ and $n-1$.
The $|r\rangle$ register, of a single qubit initialized to $|0\rangle$, stores the output of the SFC operator.

The core of the quantum procedure consists of applying Grover's search algorithm on the phase oracle, $U_{\psi}$, of the $\psi_{(x,y)}$ function, as defined in (\ref{eq:f}). 
The boolean oracle $U_{\psi}$ takes the two registers $|j\rangle$ and $|d\rangle$ as input, and is implemented through the ROT and SFC operators.
The output of the SFC operator is stored in the qubit $|r\rangle$, while the output of $U_{\psi}$ is stored on the $|out\rangle$ register,  which is initialized to $|-\rangle$ in order to make $U_{\psi}$ to behave as a phase oracle.

\begin{figure}[!t]
$$
\begin{scriptsize}
\begin{array}{lllllll}
	|j\rangle &|x\rangle &|y\rangle &|d\rangle &|r\rangle &|out\rangle & \\[0.1cm]
\hline\\
	|j\rangle &|x\rangle &|y\rangle &|d\rangle &|0\rangle &|0\rangle & ~~~~\leftarrow \textsc{initialization} \\
	|j\rangle &|\vv{x}^j\rangle &|y\rangle &|d\rangle &|0\rangle &|0\rangle & ~~~~\leftarrow \textsc{application of ROT}|j\rangle |x\rangle \\
	|j\rangle &|\vv{x}^j\rangle &|y\rangle &|d\rangle &|\psi_{(\vv{x}^j,y)}(0,d)\rangle &|0\rangle & ~~~~\leftarrow \textsc{application of SFC}|x\rangle |y\rangle |d\rangle |r\rangle \\
	|j\rangle &|\vv{x}^j\rangle &|y\rangle &|d\rangle &|\psi_{(\vv{x}^j,y)}(0,d)\rangle &|\psi_{(\vv{x}^j,y)}(0,d)\rangle & ~~~~\leftarrow \textsc{application of CX}|r\rangle |out\rangle \\
	|j\rangle &|\vv{x}^j\rangle &|y\rangle &|d\rangle &|0\rangle &|\psi_{(\vv{x}^j,y)}(0,d)\rangle &  ~~~~\leftarrow \textsc{application of SFC}^{\dag}|x\rangle |y\rangle |d\rangle |r\rangle \\
	|j\rangle &|x\rangle &|y\rangle &|d\rangle &|0\rangle &|\psi_{(\vv{x}^j,y)}(0,d)\rangle & ~~~~\leftarrow  \textsc{application of ROT}^{\dag}|j\rangle |x\rangle
\end{array}
\end{scriptsize}\\[-0.3cm]
$$
\caption{\label{fig:schema_psi}The evolution of the $6$ registers ($|j\rangle |x\rangle |y\rangle |d\rangle |r\rangle$ and $|out\rangle$) involved in the computation of the boolean oracle $U_{\psi}$, whose circuit is depicted in Fig.\ref{fig:circuit_lcs}}
\end{figure}

In Fig.\ref{fig:schema_psi}, we show the evolution of the $6$ registers involved in the computation of the boolean oracle $U_{\psi}$, namely $|j\rangle |x\rangle |y\rangle |d\rangle |r\rangle$ and  $|out\rangle$ (see also Fig.\ref{fig:circuit_lcs}). 

Specifically, the application on the register $|x\rangle$ of the ROT operator, controlled by the register $|j\rangle$, allows $|x\rangle$ to be modified so that it contains the superposition of all its possible cyclic rotations. 
Next, the application of the SFC operator on the registers $|d\rangle$, $|x\rangle$, and $|y\rangle$ allows the procedure to identify a possible position $i$ (if any) for which $\vv{x}^j[i \,..\, i+d-1] = y[i\,..\,i+d-1]$. Note that the application of the SFC operator is done in parallel, for all possible rotations of the register $|x\rangle$. 
The oracle completes its computation by saving the output of the SFC operator into the $|out\rangle$ register and uncomputing the entire process by applying the inverse operators in their reverse order.

Regarding the complexity of the circuit for the search phase, we can observe that the ROT and the SFC operators have a depth equal to $\mathcal{O}(\log^2(n))$ and $\mathcal{O}(\log^3(n))$, respectively. The same is true for their inverse, while the Grover's diffuser is executed in $\mathcal{O}(\log(\log(n)))$ time. Since Grover's search requires iterating the phase oracle and diffuser a number of times equal to $\mathcal{O}(\sqrt{n})$, we can conclude that the total complexity of the search phase is equal to $\mathcal{O}(\sqrt{n}\log^3(n))$.


\smallskip

Once the value $j$ has been returned by the search phase, the circuit for the verification phase again uses Grover's search algorithm to identify the position $i$, with $0\leq i<n$, for which $\vv{x}^j[i \,..\, i+d-1] = y[i \,..\, i+d-1]$ holds. 

The circuit,  depicted in the middle of of Fig.\ref{fig:circuit_lcs}, uses the $|j\rangle$ register containing the output of the search phase, and the $|i\rangle$ register, holding the position values of the two strings, initialized to $|+\rangle^{\log(n)}$, in order to maintain, at the initial stage, the superposition of all possible position values between $0$ and $n-1$.
The qubit $|r\rangle$, initialized to $|0\rangle$, stores the output of the FPM operator.

At the heart of the quantum circuit lies the application of Grover's search algorithm to the function $\varphi_{(x,y,d)}$, as outlined in equation (\ref{eq:t}). The quantum phase oracle, $U_{\varphi}$, for the function $\varphi_{(x,y,d)}$ operates on the input register $|i\rangle$ and is executed using the ROT and the FPM operators. The output from the FPM operator is store in the register $|r\rangle$, while the output from the oracle $U_{\varphi}$ gets stored in the $|out\rangle$ register, a single qubit that is initially set to $|-\rangle$ in order to make $U_{\varphi}$ to behave as a phase oracle within the Grover's search procedure.

In Fig.\ref{fig:schema_varphi} we schematize the evolution of the $7$ registers ($|i\rangle |j\rangle |x\rangle |y\rangle |d\rangle |r\rangle |out\rangle$) involved in the computation of the boolean oracle $U_{\varphi}$, as depicted in Fig.\ref{fig:circuit_lcs}.

Specifically, we apply the ROT operator on the register $|x\rangle$, controlled by the register $|j\rangle$, allowing $|x\rangle$ to be modified in order to contain the cyclic rotation of $j$ positions. 
Next, an application of the rotation operator controlled by register $|i\rangle$ to both registers $|x\rangle$ and $|y\rangle$ allows the two strings to be rotated by a shift of the same value. Note that, after the application of these operators, the register $|x\rangle$ contains the superposition of all possible rotations of $\vv{x}^j$, while $|y\rangle$ contains the superposition of all its possible rotations. Formally
the application of the FPM operator on the registers $|x\rangle$ and $|y\rangle$ allows the procedure to check if $\vv{x}^j[0 \,..\, d-1] = y[0\,..\,d-1]$, for all possible rotations of $\vv{x}^j$ and $y$.
The oracle completes its computation by saving the output into the $|out\rangle$ register and uncomputing the entire process by applying the inverse operators in reverse order.

\begin{figure}[!t]
$$
\begin{scriptsize}
\begin{array}{llllllll}
	|i\rangle &|j\rangle &|x\rangle &|y\rangle &|d\rangle &|r\rangle &|out\rangle & \\
\hline\\
	|i\rangle &|j\rangle &|x\rangle &|y\rangle &|d\rangle &|0\rangle &|0\rangle &  \textsc{initialization} \\
	|i\rangle &|j\rangle &|\vv{x}^j\rangle &|y\rangle &|d\rangle &|0\rangle &|0\rangle &  \textsc{application of ROT}|j\rangle |x\rangle \\
	|i\rangle &|j\rangle &|\vv{x}^{j+i}\rangle &|y\rangle &|d\rangle &|0\rangle &|0\rangle &  \textsc{application of ROT}|i\rangle |x\rangle \\
	|i\rangle &|j\rangle &|\vv{x}^{j+i}\rangle &|\vv{y}^i\rangle &|d\rangle &|0\rangle &|0\rangle &  \textsc{application of ROT}|i\rangle |y\rangle \\
	|i\rangle &|j\rangle &|\vv{x}^{j+i}\rangle &|\vv{y}^i\rangle &|d\rangle &|\varphi_{(\vv{x}^{j+i},\vv{y}^i)}(0,0,d)\rangle &|0\rangle &  \textsc{application of FPM}|x\rangle |y\rangle |d\rangle |r\rangle \\
	|i\rangle &|j\rangle &|\vv{x}^{j+i}\rangle &|\vv{y}^i\rangle &|d\rangle &|\varphi_{(\vv{x}^{j+i},\vv{y}^i)}(0,0,d)\rangle &|\varphi_{(\vv{x}^{j+i},\vv{y}^i)}(0,0,d)\rangle & \textsc{application of CX}|r\rangle |out\rangle \\
	|i\rangle &|j\rangle &|\vv{x}^{j+i}\rangle &|\vv{y}^i\rangle &|d\rangle &|0\rangle &|\varphi_{(\vv{x}^{j+i},\vv{y}^i)}(0,0,d)\rangle &  \textsc{application of FPM}^{\dag}|x\rangle |y\rangle |d\rangle |r\rangle \\
	|i\rangle &|j\rangle &|\vv{x}^{j+i}\rangle &|y\rangle &|d\rangle &|0\rangle &|\varphi_{(\vv{x}^{j+i},\vv{y}^i)}(0,0,d)\rangle &  \textsc{application of ROT}^{\dag}|i\rangle |y\rangle \\
	|i\rangle &|j\rangle &|\vv{x}^{j}\rangle &|y\rangle &|d\rangle &|0\rangle &|\varphi_{(\vv{x}^{j+i},\vv{y}^i)}(0,0,d)\rangle &  \textsc{application of ROT}^{\dag}|i\rangle |x\rangle \\
	|i\rangle &|j\rangle &|x\rangle &|y\rangle &|d\rangle &|0\rangle &|\varphi_{(\vv{x}^{j+i},\vv{y}^i)}(0,0,d)\rangle &   \textsc{application of ROT}^{\dag}|j\rangle |x\rangle
\end{array}
\end{scriptsize}\\[-0.4cm]
$$
\caption{\label{fig:schema_varphi}The evolution of the $7$ registers ($|i\rangle |j\rangle |x\rangle |y\rangle |d\rangle |r\rangle |out\rangle$) involved in the computation of the boolean oracle $U_{\varphi}$, as shown in Fig.\ref{fig:circuit_lcs}}
\end{figure}

Regarding the complexity of the circuit for the verification phase, we observe that three ROT operators have a depth equal to $\mathcal{O}(\log^2(n))$, as well as their inverse.  The FPM operator is executed in $\mathcal{O}(\log^3(n))$ time, while the Grover's diffuser on the register $|i\rangle$ requires $\mathcal{O}(\log(\log(n)))$ time. Since Grover's search requires $\mathcal{O}(\sqrt{n})$ iteration, we can conclude that the total complexity of the verification phase is equal to $\mathcal{O}(\sqrt{n}\log^3(n))$.

\smallskip

The circuit for the final check takes as input two registers, $|i\rangle$ and $|j\rangle$, containing the output of the search phase and the verification phase, respectively, and checks whether $\vv{x}^j[i \,..\, i+d-1] = y[i \,..\, i+d-1]$.
Such circuit is obtained by means of the boolean oracle $U_{\varphi}$. 
Thus, the resulting circuit for the final check requires a $\mathcal{O}(\log^3(n))$ running time to be executed.

Ultimately, the three phases achieve $\mathcal{O}(\sqrt{n}\log^3(n))$, $\mathcal{O}(\sqrt{n}\log^3(n))$, and $\mathcal{O}(\log^3(n))$ time complexity, respectively. This allows us to state that the quantum iterative test has a $\mathcal{O}(\sqrt{n}\log^3(n))$ overall time complexity and that, therefore, the \textsc{Quantum-LCS} algorithm admits an effective implementation that achieves a $\mathcal{O}(\sqrt{n}\log^4(n))$ time complexity in the case of binary strings. This complexity grows by a logarithmic factor, reaching $\mathcal{O}(\sqrt{n}\log^5(n))$ in the general case.

\section{Computing LPS and Final Discussion}\label{sec:lps}

In this paper we presented the first quantum algorithm for the LCS running in $\tilde{\mathcal{O}}(\sqrt{n})$ time in the circuit-based computational model, where $n$ is the length of the two input strings. 
We~also provided an effective implementation of the algorithm based on the circuit model and achieving $\mathcal{O}(\sqrt{n}\log^4(n))$ time complexity.

Although our discussion has focused, almost exclusively, on providing an algorithm for the LCS problem, our solution can be easily adapted for solving the LPS problem with the same complexity.
For lack of space, here we limit ourselves to describe the algorithm for computing the LPS in a very abstract way, and we reserve to describe the technical aspects of such a solution in more detail in our future works or in an extended version of this paper.

It is also a quantum-classical hybrid algorithm. The main loop, based on classical computation, consists of an iterative binary search for the length of the LPS, while the quantum part consists in the iterative test.
Since the LPS problem is computationally simpler than that for LCS, the iterative quantum test consists of only two phases, namely the search phase and the final check.

The quantum procedures that implement the two phases use a single phase oracle $U_{\rho}$, defined on the function $\rho \colon \{0,\ldots,n-1\} \times \{0,\ldots,n-1\} \rightarrow \{0,1\}$, where   $\rho_{(x,y)}(i,d) = 1$  if $\vv{x}^j[i \,..\, i+d-1]$ is palindromic, and $\rho_{(x,y)}(i,d) = 0$  otherwise.
Therefore, the oracle $P_{\varphi}$ operates so as to achieve the phase transformation 
$
	P_{\rho} |i \rangle |d \rangle  = (-1)^{\rho(i,d)} |i \rangle |d \rangle
$, for all $i,d \in \{0,1\}^{\log(n)}$.

Specifically, the search phase is responsible to identify a position $i$ (if any) such that  $x[i \,..\, i+d-1]$ is palindromic. Such phase is based on a Grover search that iterates $\mathcal{O}(\sqrt{n})$ times the oracle $U_{\rho}$.
Once the index $i$ has been identified in the search phase, the final check verifies whether $x[i \,..\, i+d-1]$ is palindromic, through a single execution of $U_{\rho}$.
The iterative test thus requires $\mathcal{O}(\sqrt{n})$ queries to the oracle. Since the binary search requires $\mathcal{O}(\log(n))$ iterations, the algorithm for computing the LPS requires $\mathcal{O}(\sqrt{n}\log(n))$ oracle queries.

Note that the oracle $U_{\rho}$ can be easily implemented through a rotation operator applied on $x$ and an inverse match operator that compares the prefix of $x$ with length $d$ against its reverse. Such an operator can be easily implemented with a depth of $\mathcal{O}(\log^2(n))$. Therefore, it follows that the algorithm for LPS admits an effective implementation running in $\mathcal{O}(\sqrt{n}\log^3(n))$ time for the case of binary strings. 
A logarithmic factor appears in the general case, obtaining a $\mathcal{O}(\sqrt{n}\log^4(n))$ time complexity.

\newpage
\bibliographystyle{plain}

\bibliography{quantum}






\end{document}